\documentclass[
aps,
prl,
reprint,
superscriptaddress,
amsmath,
amssymb,
]{revtex4-1}

\usepackage{times}
\usepackage{microtype}
\usepackage{graphicx}
\usepackage{upgreek}
\usepackage{bm}
\usepackage{placeins}

\usepackage{hyperref}
\hypersetup{
    colorlinks=true,       
    linkcolor=black,       
    citecolor=black,       
    filecolor=black,       
    urlcolor=black,        
    menucolor=black,       
}

\usepackage[
paperwidth=210mm,
paperheight=276mm,
centering,
height=24cm,
width=18cm]{geometry}

\renewcommand\thetable{\arabic{table}}

\renewcommand{\pi}{\uppi}

\newcommand*{\abs}[1]{\ensuremath{\lvert #1 \rvert}}
\newcommand*{\norm}[1]{\ensuremath{\lVert #1 \rVert}}
\newcommand*{\Abs}[1]{\ensuremath{\left\lvert #1 \right\rvert}}
\newcommand*{\Norm}[1]{\ensuremath{\left\lVert #1 \right\rVert}}

\renewcommand*{\vec}[1]{\ensuremath{\mathbf{#1}}}

\DeclareMathOperator*{\avg}{avg}
\DeclareMathOperator*{\argmax}{arg\,max}

\DeclareMathOperator{\dist}{dist}

\DeclareMathOperator{\re}{Re}

\newcommand{\VV}{\mathbb{V}}

\hypersetup{
	pdfauthor={Michael Herold, Earl T. Campbell, Jens Eisert, Michael J. Kastoryano},
	pdftitle={Cellular-automaton decoders for topological quantum memories}
}

\begin{document}
\title{Cellular-automaton decoders for topological quantum memories}

\author{Michael\ Herold}
\email[m.herold@fu-berlin.de]{}
\affiliation{Dahlem Center for Complex Quantum Systems, Freie Universit{\"a}t Berlin, 14195 Berlin, Germany}

\author{Earl\ T.\ Campbell}
\affiliation{Dahlem Center for Complex Quantum Systems, Freie Universit{\"a}t Berlin, 14195 Berlin, Germany}
\affiliation{Department of Physics and Astronomy,
University of Sheffield, Sheffield S3 7RH, UK}

\author{Jens\ Eisert}
\affiliation{Dahlem Center for Complex Quantum Systems, Freie Universit{\"a}t Berlin, 14195 Berlin, Germany}

\author{Michael\ J.\ Kastoryano}
\affiliation{Dahlem Center for Complex Quantum Systems, Freie Universit{\"a}t Berlin, 14195 Berlin, Germany}
\affiliation{Niels Bohr Institute, University of Copenhagen, 2100 Copenhagen, Denmark}

\begin{abstract}
We introduce a new framework for constructing topological quantum memories, by recasting error recovery as a dynamical process on a field generating cellular automaton. We envisage quantum systems controlled by a classical hardware composed of small local memories, communicating with neighbours, and repeatedly performing identical simple update rules. This approach does not require any global operations or complex decoding algorithms. Our cellular automata draw inspiration from classical field theories, with a Coulomb-like potential naturally emerging from the local dynamics. For a 3D automaton coupled to a 2D toric code, we present evidence of an error correction threshold above $6.1\%$ for uncorrelated noise. A 2D automaton equipped with a more complex update rule yields a threshold above $8.2 \%$. Our framework provides decisive new tools in the quest for realising a passive dissipative quantum memory. 
\end{abstract}

\maketitle  

\section{Introduction}

Prolonging the lifetime of quantum information stored in a quantum device is a monumental challenge. Yet, it is the necessary first step in the effort to scale up 
quantum computing and quantum communication to a commercially viable level.  As quantum coherence is intrinsically fragile, it is clear that increased robustness of the encoded information needs to rely heavily on quantum error correction~\cite{Aharonov,G01a}. Topological codes in particular have emerged as the most promising quantum error correcting codes~\cite{SurfaceCode}, where the toric code 
\cite{tor,Kitaev}
is a 
paradigmatic example. However, in three or fewer dimensions, excitations propagate at little energy cost under thermal dynamics, rapidly corrupting the encoded information~\cite{yoshida2011feasibility}. Schemes based on sequential measurements of the system's error syndromes, and subsequent elimination of errors have been suggested to preserve the logical subspace.

When performing active error correction, it is essential for the decoding process to be much quicker than the decoherence time. Prior proposals have focused on the development of efficient decoding algorithms~\cite{WangFowlerHollenberg,Duclos-Cianci10,BravyiHaah} or decoders with high error thresholds \cite{Wootton12,Hutter14}. Some of them offer the possibility of parallelization enabling a runtime logarithmic in the system size~\cite{fowler2013minimum,Duclos-Cianci10}. This improves prospects, but still requires a computer with communication between many spatially separated cores.  By incorporating a message routing system into a lattice of cores, such long range communication can be achieved with only nearest neighbour connections.  However, the complexity, time lag, and communications traffic of such a system is not fully understood and has not been simulated.  Hence, key obstacles remain.

\begin{figure}[t!]
\includegraphics[width=\columnwidth]{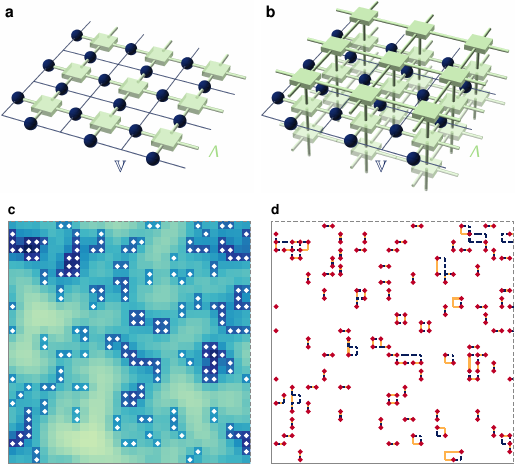}
\caption{\textbf{Schematic presentation of the cellular automaton decoders.} (\textbf{a}) The toric code on a periodic lattice controlled by a classical cellular automaton decoder on a 2D lattice. Blue spheres represent the physical qubits and the green boxes represent the elementary cells of the automaton. The communication between cells and with the syndrome measurements (anyons) is nearest neighbour, indicated via connecting tubes. (\textbf{b}) A 3D cellular automaton decoder with the central layer hosting the toric code.  (\textbf{c}) The field, encoded in the cellular automaton, generated by the anyons of the toric code. Regions with a larger density of anyons have a larger field value. The profile of the field will depend on the specific characteristics of the automaton dynamics. The cellular automaton performs two tasks: (i) it updates the field in order to propagate information between distant anyons, and (ii) it moves the anyons in the direction of largest field gradient. (\textbf{d}) Logical information associated with one decoding run. The red diamonds are the ends of error strings, the orange lines are the actual physical error lines, and the blue dotted lines are the recovery paths dictated by the decoder.}
\label{Fig1}
\end{figure}

In this work, we propose an entirely new approach towards designing topological quantum memories that naturally incorporates parallelization. Our design can be implemented via classical hardware composed of small local memories and a small set of local operations only depending on neighbouring memories and neighbouring physical qubits, without constituting universal local processors~\cite{Gacs,Harrington2004} or demanding an explicit message routing system. That is, we develop simple cellular automata which efficiently perform active error correction on the toric code. The operational principles of our automata are based on the mediation of attractive long-range interactions between excitations. That way excitations will tend to collapse together rather than to extend out and create logical errors in the code. Inspired by classical electrostatics and gravitational fields, we identify local update rules that induce such error correcting long range fields.

We provide extensive numerical analysis of three specific decoders, two of which use a 2D automaton, and the third using a 3D automaton.  Two of our decoders have comparable threshold values to more conventional decoders that require more sophisticated hardware.  We also provide detailed analysis of the equilibration and propagation characteristics of the fields generated by the automata, which justifies the choice of tunable parameters in our decoders. Finally, we study a class of long range fields and their decoding capabilities, and observe that certain fields are too long range to reliably identify excitation pairs. 

Our schemes share some features with other proposals which add an auxiliary system in order to enforce suppression of error creation~\cite{Hamma,Chesi,Pedrocchi11, Pedrocchi13,Fujii14,SelfCorrectingMemoryBoundary, ThermallyStableSurfaceCode,Weakly}.  Prior proposals typically use quantum
auxiliary systems 
instead of classical auxiliary systems for studies of self-correcting memories, where full numerical investigations are prohibitively difficult. Our class of models in some sense combines the benefits of active and of passive memories. Indeed, provided that the cellular automaton update rules  do not change in time, the decoding process can be interpolated down to continuous time. In this setting error correction acts simultaneously with error creation, effectively preventing the creating of long error strings. This would amount to a dissipative self-correcting memory~\cite{Pastawski11,DissipativeGadgets}. In this work, we focus on decoding devices but it should be clear that our design incentives promise a generalization to new schemes of dissipative self-correction. 

\section{Methods}

The main purpose of this work
is to present a fundamentally new class of decoders for the 2D toric code.  The toric code has the important feature  that it can be considered as a physical model with purely local interactions. In this model, the physical spins reside on the edges of a periodic $L\times L$ 
lattice which we henceforth denote $\VV$, and the stabilizer operators act on the four neighbouring spins in the star or plaquette formation (see Supp. Mat. for a brief review of the toric code). Errors in the toric code appear as strings, where only the endpoints of the strings are revealed in the error syndrome measurements.  We will refer to these endpoints as \emph{anyons}, and they play the role of excitations in a classical statistical model \cite{SurfaceCode,Bombin12}.  An error configuration is successfully decoded if all anyons are paired, such that all anyon paths are  contractible loops on the torus. Throughout, we assume independent single site Pauli-spin $X$ and $Z$ errors at a rate $p$. In this way both error sectors decouple, and we can treat them independently. Here we consider only  $X$ errors, the $Z$ errors can be treated identically.

The decoders we propose were conceived with the following goal in mind: To arrive at
purely local update rules that are as simple as possible. By this we mean that the decoding can be implemented with simple local units of computation that depend only on nearest neighbours. The most natural model for this type of computation is a cellular automaton, which is  indeed the nature of our class of decoders. In particular, we will consider an auxiliary cellular automaton, whose purpose  is to communicate long range information between anyons, that enables local decisions to correct errors in the toric code (see Fig.~\ref{Fig1}).

The automaton extracts anyon information from the physical qubits via local stabilizer measurements at each cell of the toric code lattice $\mathbb{V}$, the resulting outcome  is stored in some local registry which we label $q_{E}(\vec{v})$.  The automaton also uses an auxiliary classical system on another lattice $\Lambda$, and at each cell $\vec{x} \in \Lambda$ and at each time $t$ it stores a real number $\phi_t(\vec{x})$. We will refer to the auxiliary system as the \emph{field} generated by the \emph{$\phi$-automaton}.  The simplest auxiliary lattice  has cells coinciding with the toric code cell (so $\Lambda = \mathbb{V}$), but we also consider larger dimensional auxiliary systems provided they include the toric code as a sub-lattice $\mathbb{V} \subset \Lambda$. Every auxiliary lattice considered herein is either a 2D torus, so $\Lambda \cong \mathbb{V}$, or a 3D torus with the $x_{3}=0$ plane coinciding with the toric code lattice.  To simplify later expressions, we extend the domain of $q_{E}$ to the whole auxiliary lattice with the understanding that anyons never leave the toric plane, so $q_{E}(\vec{x})=0$ whenever $x_{3}\neq 0$.

\begin{table}[t]
\caption{\textbf{Summary of results for $\phi$-automaton decoders.} }
\label{tbl:results}

\vspace{1em}
    \begin{ruledtabular}
    \begin{tabular}{lccc}
    & 2D &  2D$^*$ & 3D \\
    \colrule
    Lattice ($\Lambda$) & $L \times L$ & $L \times L$ & $L \times L \times L$  \\
    Field velocity ($c$) &  $\tau $ indep. & $ 1 + 0.2\cdot \tau $ & $10\cdot\log^2 (L)$ \\
    Threshold ($p_{\text{th}}$) & N/A & $ 8.2\,\%$ & $6.1\,\%$ \\
    Required sequences ($\tau_\text{RT}$)& N/A & $o(\log^{2.5} (L) )$ & $o(\log (L))$ 
    \end{tabular}
    \end{ruledtabular}

\raggedright\footnotesize
\smallskip

$\tau$ refers to the sequence index, where each sequence contains $c+1$ 
elementary updates per cell. The decoder terminates after $\tau_\text{RT}$ sequences on average. For the 2D$^*$ decoder the rules for the field updates can be thought of as 
time dependent since $c$ increases with $\tau$. The smoothing parameter as defined in Eq.~\eqref{eq:PhiUpdate-Recursive} is $\eta=1/2$.

\end{table}

The  system's configuration at any given time 
is the triple $\{ E, q_{E}, \phi \}$, where $E$ is the actual error configuration. Assuming ideal measurements, $q_{E}$ is redundant as it is determined by the error configuration.  The dynamics of the system are divided into update sequences, labeled by $\tau$, in which
$\{ E, q_{E}, \phi \}$ evolve.  Each update sequence is subdivided into $c+1$ elementary steps, first there are $c$ repetitions of a $\phi$-update (in which
only $\phi$ changes), followed by a single anyon update (a partial error correction where $E$ and hence also $q_{E}$ change).  Since lattice cells only interact with their nearest neighbours, these elementary update rules must be local.  In a single $\phi$-update from time $t$ to $t+1$, the field $\phi_{t+1}(\vec{x})$ can only depend on $\phi_{t}(\vec{x})$, $q_{E}(\vec{x})$ and values $\phi_{t}(\vec{y})$ where $\vec{y}$ is a neighbour of $\vec{x}$.  For each excited cell (where $q_{E}(\vec{x})=1$) the anyon update rule decides whether the anyon hops to an adjacent site, and being similarly local, this rule depends only on $\phi$ values at adjacent sites.  With good design of update rules, the $\phi$ values will meditate information between anyons about the location of their closest potential partners. Later, we derive suitable updates rules, which can be roughly characterised as: take the average of your neighbours and add the charge. This corresponds to a local discretisation of  Gauss' law. 

To fuse nearby anyons we desire that each anyon moves towards regions of higher field intensity. A natural way to achieve this is with the anyon update rule:  find the adjacent cell with (unique) largest field value, then move there with probability $1/2$.  We also considered alternative anyon updates,  but none yielded higher thresholds than this simple and intuitive strategy. Physically, an anyon hopping from $\vec{x}$ to  $\vec{y}$ is implemented by a Pauli $X$ applied to the qubit on the lattice edge separating cells $\vec{x}$ and $\vec{y}$.  All anyon hops occur in parallel resulting in partial correction of the error string $E$, to a new string
$E'$ and the anyon distribution similarly updates to $q_{E'}$.  This process only removes anyons, with two anyons on the same site canceling, eventually removing all anyons and so completing a whole error correction cycle. The ratio between the number of $\phi$-updates per anyon update is labeled $c$, and can be understood as the speed of propagation of the field. We will call $c$ the \textit{field velocity}. A pseudo code of our algorithm is provided in the Supp. Mat.

\section{Results}

\begin{figure}
\includegraphics[width=\columnwidth]{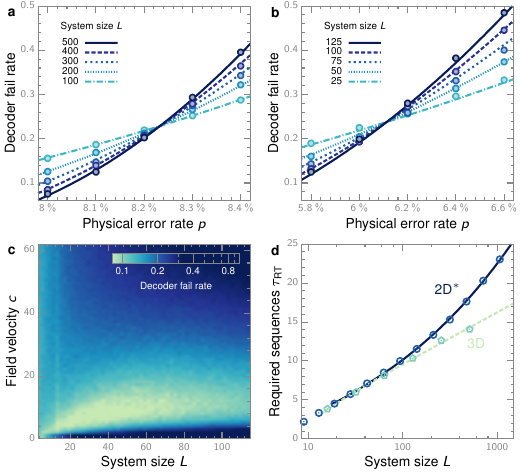}

    \caption{\textbf{Performance of $\phi$-automaton decoders.} (\textbf{a})~Threshold plot for the 2D$^*$ and (\textbf{b})~the 3D $\phi$-automaton decoder. The decoder fail rate decreases with system size below threshold and increases above threshold. The number of samples in (\textbf{a}) ranges from $2\,800$ samples per point for larger system sizes to $21\,000$ samples for smaller system sizes. In \textbf{b} the sample size per point is at least $20\,000$. The points were fitted using a universal scaling law \cite{Watson2013}. (\textbf{c})~2D decoder fail rate as a function of lattice size $L$ and field velocity $c$ for $p=6\,\%$. For very small $c$ the decoder has a large fail rates due to self interaction (see corresponding paragraph for details). At large $c$, the decoder also fails because of a stationary field with excessively long range. There exists a sweet spot around $c\in[5,15]$ and $L\in[20,80]$. (\textbf{d})~Average number of sequences required for a 2D$^*$ and 3D $\phi$-automaton decoder to remove all anyons with $p=1\,\%$. The lines represent best fits using a $\log^\beta (L)$ function for the 2D$^*$ decoder and $\log (L)$ for the 3D decoder. The optimal fit yields $\beta\approx 2.5$.}
\label{Fig_Main_Results}
\end{figure}

\medskip\noindent\textbf{Numerical results.} We numerically benchmarked our decoders using uncorrelated $X$ and $Z$ flip  noise with error probability $p$.
We say that a decoder is \emph{asymptotically-working} if it exhibits an error correction threshold $p_\text{th}$. That is to say, if $p<p_\text{th}$ then increasing the lattice size linearly exponentially suppresses the probability of the decoder creating a logical error by mismatching anyons.  Numerically, the threshold appears as a crossing for various lattice sizes, such as in Fig.~\ref{Fig_Main_Results}a,b.  Further numerical results showing details of error suppression are provided in the supplementary material.

Our results are summarised in Table 1.  We present three decoders called the 2D, 2D$^*$ and 3D automata.  The 2D$^*$ and 3D cases yield asymptotically working decoders, whereas the simple 2D decoder only suppresses errors up to a finite size lattice (e.g. $L \sim 60$, see Fig.~\ref{Fig_Main_Results}c).  While the 2D decoder is expected to be most
suitable for modest storage needs, no strict asymptotic threshold exists.  We consider two more decoders using precisely the same update rules, but different only either in how these updates are composed or in the lattice they are implemented on. The 2D$^*$ $\phi$-automaton keeps the 2D lattice, but alters how updates are composed by letting $c$ increase with each sequence $\tau$, which increases the ratio of  field updates to anyon updates as time increases. We find the  2D$^*$ $\phi$-automaton has a threshold of $\sim 8.2\%$ (see Fig.~\ref{Fig_Main_Results}a). Our second variant is the 3D $\phi$-automaton, which keeps $c$ constant in time, but implements the field updates on a 3D lattice.  The toric code remains 2D, but is embedded in a plane of the 3D $\phi$-automaton (see Fig.~\ref{Fig1}b).  The 3D $\phi$-automaton exhibits a noise threshold at $\sim 6.1\%$ (see Fig.~\ref{Fig_Main_Results}b). Although $c$ is constant in time, it must 
scale logarithmically with $L$. These thresholds are only a few percent below the best thresholds using a centralized computing architecture. Further numerical data on performance below threshold is presented in the supplementary material.

Ideally, we would like all update rules to be completely independent of the system size $L$, and not to change throughout sequences. Yet, we find that this is not possible within the framework of $\phi$ automaton decoders. Indeed, simulations show that in order for the decoder to converge in a time which is sub exponential in $L$, the field has to have propagated a distance of order $\log (L)$, which corresponds to a lower bound on the field velocity $c \geq c_{\text{min}} \sim \log^2(L)$. The dependence on $\log(L)$ can be understood by the fact that maximal clusters of errors below the error threshold are of order $\log(L)$ (see Refs.~\cite{kovalev2013fault,gottesman2013overhead} and Supp. Mat.), and the decoder must collapse these maximal clusters in a time proportional to their size. If $c$ were to be taken independent of $L$, then beyond a critical cluster size, the field contribution due to an anyon's self-interaction would dominate the contribution from the anyons at the other end of the cluster. Self-interaction prevents constant $c$ decoders from converging in a time polylogarithmic in $L$, and the phenomena is discussed in detail later.  Thus, in any field based model, the field velocity must scale with the system size. Later in the paper, we give an explicit lower bound on $c$ which is derived from the equilibration properties of the $\phi$-automaton. 

Given that we want our class of decoders to be adaptable to a setting where measurement data is regularly refreshed, it is desirable to have update rules that are invariant throughout sequences. 
However, this appears not to be possible when the fields are restricted to two spatial
dimensions (see Fig.~\ref{Fig_Main_Results}c). The $\phi$-automaton decoder works asymptotically in three or higher dimensions because of the steady state profile of the field. Indeed, the steady state poisson field of a single charge scales as $\log(r)$ in two dimensions and as $r^{2-D}$ in higher dimension, where $r$ is the distance from the source charge. From numerical simulations, we are lead to conclude that the 2D equilibrium field is too long range, and tends to break the cluster structure of the anyons by extending error strings rather than shrinking them.  In higher dimensions the field profile decays steeply enough so that this is not the case. We also later address whether there is a proper threshold between the $\log(r)$ and the $1/r$  decaying fields.

Should we not be concerned with the field velocity changing across sequences, the 2D$^*$ $\phi$-automaton decoder is preferable as it exhibits a higher threshold and can more easily be implemented in an integrated circuit type architecture \cite{Martinis1,barends2014}. With $c$ increasing linearly at each update sequence, this allows
elements within the clusters to pair up, while simultaneously preventing the field from extending across clusters. The main mechanism  responsible for the success of the 2D$^*$ decoder  is analogous to the ``Expanding Diamonds" decoder from Refs.~\cite{DennisPhD,Wootton}. 

Finally, in Fig.~\ref{Fig_Main_Results}d we summarise  the runtime estimates of our two working $\phi$-automaton decoders. The 3D decoder completes an error correction cycle, pairing all anyons, in the order of $\log(L)$ update sequences, while the 2D$^*$ decoder terminates after $\sim \log^{2.5}(L)$ sequences. This can be seen as very strong evidence that the typical maximal error cluster is indeed of size $\sim\log(L)$, and that our decoders collapse these clusters in optimal time. The 2D$^*$ decoder has a runtime $\sim\log^{2.5}(L)$ sequences as the value of $c$ needs to exceed $\sim \log^2(L)$ before the $\phi$-automaton can transmit information across a maximal cluster to collapse it. It should be noted that our estimate of the exponent of $2.5$ is based on fitting a polylogarithmic function using 11 data points, and could be slightly off. Finally, each sequence is composed of $c$ field updates and one anyon update, and so we can also quantify runtime  in units of updates.  For the 3D decoder, $\log(L)$ sequences with $c =10\cdot \log^2(L)$ gives a runtime of order $\log^3(L)$ updates.  For the 2D decoder $\log^{2.5}(L)$ sequences with $c = 1+\tau/5$, we sum over $\tau=1,\dots,\log^{2.5}(L)$ to get a runtime of order $ \log^5(L)$ updates. Note that for the 3D decoder the runtime implies that asymptotically the vast majority of cells in the third dimension is not reached by any field information. Therefore the size of the third dimension has to scale as $\log^3 (L)$ only.

\medskip\noindent\textbf{The $\phi$-automaton.} In this section we explore in detail the local update rules of the $\phi$-automaton from physically motivated considerations. The key idea is to emulate a long-range field, such as from Newton's law of gravitation or Coulomb's law, which mediates an attractive force between the excitations of the toric code. 
In nature such long range forces emerge from simple local dynamics in the field. Here we  achieve the same effect on a discrete lattice governed by a cellular automaton. The notion of mutually attracting particles with equal charge rather corresponds to gravitational fields, but we will stick to a language that is mainly inspired by electrodynamics. 

To elaborate on this analogy, we detour into electrostatics, where the electric field is the gradient of a scalar potential $\Phi$. Gauss's law simplifies to Poisson's equation
\begin{equation}
    \nabla^2 \Phi(\vec{x}) = \sum_{j=1}^{D} \frac{ d^2 \Phi(\vec{x}) }{d x_{j}^2} = q(\vec{x}) ,
\end{equation}
where $D$ is the spatial dimension and $q$ is the charge distribution.  The only isotropic solution of Poisson's equation with a single charge at the origin is
\begin{equation}
\label{ContVar}
\Phi(\vec{x}) = \begin{cases}
-\log r & {\rm for } ~D=2 ,\\
r^{2-D} & \text{otherwise,}
\end{cases}
\end{equation}
where $r=\dist(\vec{x},\vec{0})$
denotes the distance of some lattice site $\vec{x}$ from the source charge at the origin. The minus in front of the logarithm is chosen to ensure that $\Phi$ is convex for all $D$. The variety of long range behavior motivates the use of $\Phi$ as an information mediator. Our goal is now to approximate $\Phi$ via a cellular automaton.

A convenient discretization of the derivative is $\operatorname{d}: \phi(x) \mapsto \phi(x+1/2)-\phi(x-1/2)$. Applying this prescription twice yields the double derivative $\operatorname{d}^2: \phi(x) \mapsto \phi(x+1)+\phi(x-1)-2\phi(x)$. On a $D$-dimensional square lattice $\Lambda$, this generalizes to the discrete Laplacian operator
\begin{equation}\label{eq:discr_poisson}
   \nabla^2 \phi(\vec{x}) = -2 D \phi(\vec{x}) + \sum_{ \langle \vec{y}, \vec{x} \rangle} \phi(\vec{y}) ,
\end{equation}
where the sum $  \langle \vec{y}, \vec{x} \rangle$
is over all cells $\vec{y}$ neighbouring cell $\vec{x}$; so all $\vec{y}$ for which $\dist(\vec{x},\vec{y})=1$.

We proceed by specifying a set of dynamical equations, or $\phi$-automata update rules, whose stationary distribution satisfy the discrete Poisson equation $\nabla^2 \phi(\vec{x}) = C q( \vec{x})$.  Here the role of the charge is replaced by anyonic excitations, and we identify $q$ with $q_E$. Note that only gradients of $\phi$ will be considered meaningful for anyon movement, and so fields differing by an additive constant are deemed equivalent. Invoking the Jacobi method \cite{saad2003iterative}, we consider the following $\phi$-automaton update rule
\begin{equation}
\label{eq:PhiUpdate-Recursive}
    \phi_{t+1}(\vec{x}) = (1-\eta)  \phi_{t}(\vec{x}) + \frac{\eta}{2 D}\sum_{ \langle \vec{y}, \vec{x} \rangle}  \phi_{t}(\vec{y})  +q_{E}(\vec{x})  ,   \end{equation} 
where $0 < \eta < 1/2$ is a smoothing parameter we can freely choose and for convenience the unit charge is set to $C=-2 D / (1-\eta)$. These update rules are manifestly local. We can cast Eq.~\eqref{eq:PhiUpdate-Recursive} as a matrix equation $\phi_{t+1}=G\phi_{t}+ q_E$ which should be read as
\begin{equation}
\label{EQN_update}
\vec{\phi}_{t+1}(\vec{x})=\sum_{\vec{y}} G_{\vec{x}, \vec{y}} \vec{\phi}_t(\vec{y}) + q_E(\vec{x}),
\end{equation}
where $G$ is a doubly stochastic block
circulant matrix encoding the $\phi$-automaton update steps.  This reformulation allows us to leverage the machinery of matrix analysis \cite{Bhatia}.

\medskip\noindent\textbf{Stationary states and equilibration.}  Assume for now that the anyon configuration is fixed. By recursively iterating Eq.~\eqref{EQN_update}, we get
\begin{equation}
\vec{\phi}_{t}= G^t \vec{\phi}_0 +\sum_{m=0}^{t-1}G^m q_E.
\end{equation}
It is easy to see that in general the $\phi$-automaton satisfying Eq.~\eqref{eq:PhiUpdate-Recursive} does not converge, since the total field values build up steadily due to the charges.  Indeed, when $Q=\sum_{\vec{x}}q_{E}(\vec{x})$ denotes the total charge present in the lattice, the total field increases by $Q$ with each field update.  However, it is important that the gradient of the field equilibrates towards a fixed value.  With this in mind, we define $\tilde{\phi}_{t}$ to be the rescaled version of $\phi_t$
such that $\tilde{\phi}_{t}(\vec{x})=\phi_t(\vec{x})-K$, where the constant $K$ is chosen such that $\sum_\vec{x}\tilde{\phi}_{t}(\vec{x})=0$.

The matrix $G$ can be diagonalized by Fourier transform (see the Supp. Mat.), 
allowing us to obtain a unique solution.  We find that the stationary field produced by a single anyon at the origin is
\begin{equation}\label{steady_state_field}
 \varphi(\vec{x}):= \lim_{t\to\infty} \tilde{\phi}_{t}(\vec{x}) =  L^{-D}\sum_{\vec{k} \in \Lambda, \vec{k} \neq \vec{0}} (1- \lambda_{\vec{k}})^{-1} e^{ i  \vec{k}\cdot\vec{x} },
\end{equation}
where the sum is over all Fourier components except the zero vector and $\lambda_{\vec{k}}$ are the eigenvalues of the matrix $G$,
\begin{equation}\label{Eigenvalues}
    \lambda_{\vec{k}} =  1-\eta+\frac{\eta}{D}\sum_{j=1}^{D} \cos ( k_{j} ) .
\end{equation}
The $\vec{k}=\vec{0}$ term is excluded from the sum in Eq.~\eqref{steady_state_field} which would provide an additive constant to the field. By omitting this term we fix the normalization so that $\sum_\vec{x} \varphi(\vec{x})=0$.  It is also easy to see that the stationary distribution of a charge at cell $\vec{y}$ is given by $\varphi(\vec{x} - \vec{y})$. By linearity,  the $\phi$-automaton satisfies the superposition principle of fields, and hence the stationary distribution of a collection of charges is just the sum of the distributions of each individual charge. That is, 
$\tilde{\phi}_{\infty}(\vec{x})=\sum_{\vec{y}} q_{E}(\vec{y}) \varphi(\vec{x}-\vec{y})$.  

The distance between $\tilde{\phi}_{t}$ and $\tilde{\phi}_{\infty}$ (as measured by Euclidean distance) can be shown by using matrix inequalities to decrease exponentially fast, so that (details in Supp. Mat.)
\begin{equation}
    \norm{ \tilde{\phi}_{t}-\tilde{\phi}_{\infty} }_2 \leq e^{-(\eta \pi^2 /D) t/L^2}  \norm{ \tilde{\phi}_{0}-\tilde{\phi}_{\infty} }_2 ,
\end{equation}
where the distance is measured in the vector 2-norm $\norm{\vec{u}}_2^2=\vec{u}\cdot \vec{u}$.  This would suggest a relaxation time of the $\phi$-automaton of the order of $L^2$, which can be understood as diffusive spreading of information. In the following section we will argue that the relevant contribution of the field converges in a time much faster for our decoders, because we are only interested in information propagating on distance scales of the order $\log(L)$, the maximal error cluster size. 

\begin{figure}
\includegraphics{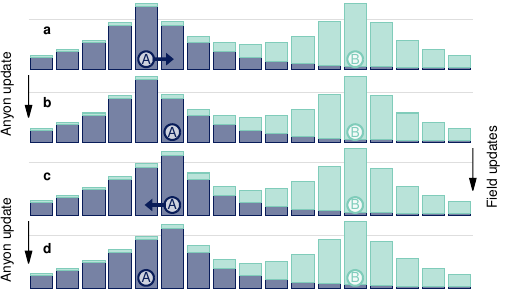}
 \caption{\textbf{Self-interaction} A setup with two isolated
 anyons $A$ and $B$, and their fields in a self-interaction cycle.  Suppose that the fields from each anyon starts in its stationary configuration (\textbf{a}). After one anyon update move, anyon $A$ moves in the direction of largest field gradient -- towards anyon $B$. (\textbf{b}) Immediately after anyon $A$ moves, the field which it leaves behind is much larger that the field from anyon $B$. (\textbf{c}) After a certain number of field updates, the residual self field from anyon $A$ decreases, but might still tell the anyon to move in the wrong direction (\textbf{d}).}
    \label{Fig_Flickering}
\end{figure}

\medskip\noindent\textbf{Self interaction.} The $\phi$-automaton carries information about distant anyons through local update rules. As has already been hinted at in the previous section, information takes time to propagate through the field, and the most relevant manifestation is in the pernicious phenomenon that we refer to as \emph{self interaction}. The field around an anyon is generated both by distant, potential partner anyons, but also by itself. This is a feature shared by all field theories. We can understand this self interaction by considering two close-by anyons $A$ and $B$, as in Fig.~\ref{Fig_Flickering}. Our story begins with anyon $A$ moving in the direction of largest local field gradient, towards anyons $B$. After anyon $A$ has moved, it leaves behind  its own field, which will typically be a local maximum of the $\phi$-automaton. In order for anyon $A$ at the end of the following sequence to again move in the direction of anyon $B$, the field gradient generated by anyon $B$ has to be larger than the residual self field gradient around the cell of anyon $A$. We call this phenomenon self interaction, and next derive conditions on the field velocity to prevent this from occurring.

Suppose anyon $A$ starts at the origin and anyon $B$ is at cell $\vec{y}$.  Assuming the initial configuration is at equilibrium, $\phi_{0}(\vec{x})=\varphi(\vec{x})+\varphi(\vec{x}-\vec{y})$.  After anyon $A$ moves, the field begins to relax towards $\phi_{\infty}(\vec{x})=\varphi(\vec{x}-\vec{e})+\varphi(\vec{x}-\vec{y})$, where $\vec{e}$ is a unit vector in the direction of anyon $B$.  We want to estimate the time required before the field gradient generated by anyon $B$ is larger than the self-interaction of anyon $A$. 
In other words,  assuming $\phi_0(\vec{x})=\varphi(\vec{x})+\varphi(\vec{x}-\vec{e})$, after what time do we have 
\begin{equation}\label{eqn:converg1}
    \abs{ \phi_{t}(\vec{x})-\phi_{\infty}(\vec{x}) }   \leq g_l,
\end{equation}
where $g_l$ is the gradient at the origin due to a charge at a distance $l=\dist(\vec{0},\vec{y})$. It is not difficult to see that the left hand side of Eq.~\eqref{eqn:converg1} does not depend on the field from  anyon $B$.

A straightforward, but lengthy, calculation (in the Supp. Mat.) shows that  for all $\epsilon >0$
\begin{equation}
    \abs{ \phi_{t}(\vec{x})-\phi_{\infty}(\vec{x}) } \leq \epsilon
\end{equation}
whenever
\begin{equation}
\label{tmin_rough}
	t \geq  \chi(\vec{x}) \epsilon^{-2/(D-1)},
\end{equation}
where $\chi$ is a function of position, but is independent of $t$ (see Supp.\ Mat. for the
exact form). Considering the second anyon at point $\vec{y}= \vec{e}$, it creates a field gradient near the first anyon (at the origin)
\begin{equation}
    \nabla \varphi(\vec{x}-\vec{y}) |_{\vec{x=0}} = 
    \sum_{j} (\varphi(-\vec{y})-\varphi(\vec{e}_{j}-\vec{y})) 
    \cdot \vec{e}_{j},
\end{equation}
where $\{\vec{e}_{j}\}$ is an orthonormal basis of unit vectors.  Therefore, we must require that
\begin{equation}
     \chi(0) \epsilon \leq   \abs{ \nabla \varphi(\vec{x}-\vec{y}) |_{\vec{x=0}}  } := g_{\vec{l}}.
\end{equation}
By percolation arguments, the maximum  cluster size of anyons can be argued to typically be order $l \sim \log ( L )$.  
Given that the gradient is of the order $l^{1-D}$, require $\epsilon \sim \log^{1-D}(L)$,
\begin{equation}
\label{EQ_cMin}
     c \geq  c_{\rm crit}\sim \log^2(L),
\end{equation}
where the dependence on $D$ has now canceled.

\medskip\noindent\textbf{The field profile.} We already discussed in detail that when the field velocity $c$ is kept invariant throughout sequences,   the 2D and 3D decoders exhibit fundamentally different behavior.  Here we supplement these results by investigating another decoder model, not a cellular automata, to investigate the large $c$ regime when fields are always close to stationary solutions.  In order to address this question, we consider a class of power law potentials

\begin{equation}
	\Phi(r) = r^{-\alpha}
\end{equation}
for $\alpha>0$.

\begin{figure}[t]
\includegraphics[width=\columnwidth]{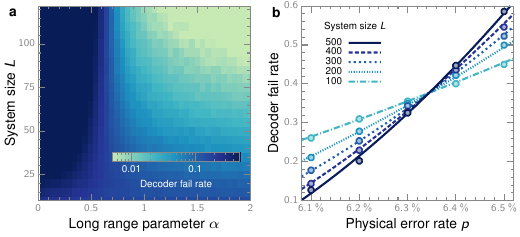}
    \caption{(\textbf{a}) The failure rate $p_{\text{fail}}$ when the initial error rate is $p=5\,\%$, for a range of lattice sizes $L$ and a range of idealised decoders parameterised by $\alpha$.  These decoders are 
    based on an auxiliary system consisting of explicit non-local computation using Coulomb
    potentials of $1/r^{\alpha}$.  At small $\alpha$, $\alpha \leq 1$, 
    the decoder is improper with large lattice sizes yielding large failure rates.  Whereas, at larger $\alpha$ error correction is observed with increasing 
    lattice size. (\textbf{b}) Threshold plot for the idealised decoder with a parameter value $\alpha=1$, which is very close to the behavior observed with the 3D $\phi$-automaton.}
        \label{Fig_Vary_Alpha}
\end{figure}

We simulate anyon movements in these \emph{perfect} fields, by inserting the sum of all anyon fields by hand, instead of simulating it locally by a cellular automaton. We alternate between these instantaneous field updates and the anyon update rule as defined before. For $\alpha=1$ we expect a behavior similar to the decoder with 3D $\phi$-automaton. 
The benchmark presented in Fig.~\ref{Fig_Vary_Alpha}b indeed shows a threshold at $p_{\text{th}}\sim6.3\,\%$, which is very close to the threshold of $\sim 6.1\,\%$ seen for the 3D $\phi$-automaton.  This independently confirms that our choice of finite $c$ is sufficiently large for equilibration between distinct anyon moves.

Seeking to understand the transition between the 2D and the 3D stationary field behavior, we consider various values of $\alpha$ and lattice sizes for fixed $p=5\,\%$ in Fig.~\ref{Fig_Vary_Alpha}a. In general the decoder fail rate reduces with increasing $\alpha$. This supports our claim that fields with a shorter range are more suited for efficient shrinking of errors. However, our main observation is that for $\alpha<0.5$ the decoders fail rate increases drastically with the system size $L$. This means that the decoder does not exhibit a threshold above $p=5\,\%$ for $\alpha<0.5$. Although it is difficult to certify numerically, for $0.5\leq\alpha\leq 0.7$ we did not either find an indication of a threshold at any $p$ up to lattice size $L=400$.

From these observations we infer that there has to exist a critical $\alpha_{T}$ such that $\Phi$ provides a asymptotically-working decoder. For values of $\alpha$ below the transition, the increased contribution from far away anyons leads to the misidentification of error strings.  Recall that anyons move according to the maximal field gradient 
of $r^{-\alpha}$,
which is proportional to $r^{-(\alpha+1)}$. The total field at a point is the sum of the contributions from all of the anyons, hence for an infinite lattice the gradient is only guaranteed to be finite for 
 $1/r^{\beta}$, when $\beta>2$. Only for $\alpha>1$ does the gradient have a well defined value asymptotically. Hence we expect $\alpha_{T}=1$ for the 2D toric code.  This intuition is further supported by a simple simulation of the 1D toric code (i.e., the repetition code), where we observe a clear transition at $\alpha=0$, which corresponds to a gradient of $1/r$. Here, we have allowed for values of $\alpha\leq 0$
 by setting $\Phi({r}) = - r^{-\alpha}$ for $\alpha<0$ and $\Phi=-\log({r})$ for $\alpha=0$.

We highlight that very high values of $\alpha$ are not favourable in general, since they require increased precision in the field  resolution. Meanwhile, we expect that the 3D $\phi$-automaton always gives rise to a field with slightly  shorter range than $1/r$ (i.e., $\alpha>1$). Therefore,
the 3D $\phi$-automaton seems to be sitting exactly at the sweet spot, providing a functioning long range $\phi$-field decoder with a maximally robust field.

\section{Discussion}

In this work, we have introduced a new  class of decoders that pair anyonic excitations by mediating long range information through an auxiliary field. The field, and the anyon movement, is generated by a cellular automaton, in such a way that the decoder has an intrinsically parallelised architecture.  We have observed that the attractive interactions mediated by the cellular automaton can stabilise topological states. Further we have found indications that  the stability exhibits a phase transition in the long-range parameter of the attractive field. We have identified two particular decoders within this class that exhibit an error threshold, one requiring a 3D auxiliary system with homogeneous update rules, with an error threshold above $6.1\,\%$, and another with a 2D auxiliary system and time dependent update rules with an error threshold above $8.2\,\%$.  When below threshold both decoders also show noise suppression reducing exponentially with lattice size.  Our schemes have two main conceptual limitations in their present formulation. First, the processing cost is determined by the field velocity $c$ that must scale polylogarithmically with the system size $L$. Second, the field needs a precision sufficient to distinguish the presence of an anyon distance $\sim \log(L)$ away. Encoding the field digitally, this is achieved by a local field register of $\log^2(L)$ bits at every field site. At the same time, it is important to emphasize once again that the communication requirements are minimal, relying on nearest-neighbour communication only, requiring no wiring or long-distance communication. Next we discuss in detail how these costs compare with other approaches, and we will see that such logarithmic costs are generic.

The first, and most common approach to decoding errors is primarily designed for a serial computing architecture.  Compared to these proposals our threshold values are only modestly smaller than the best decoders using Monte Carlo techniques~\cite{Wootton12,Hutter14} or the minimum weight perfect matching (MWPM) algorithm \cite{fowler09-2}, but are comparable  to recent popular proposals based on real-space renormalization techniques~\cite{BravyiHaah,Duclos-Cianci10} or so-called expanding diamonds \cite{DennisPhD,Wootton}. However, our cellular automata decoders have a strict locality neighbourhood, and are hence potentially very attractive for implementations of  topological quantum memories in integrated circuit type architectures, as were recently proposed in Refs. \cite{Martinis1,barends2014}.  Serially designed decoders allow for some parallelisation, and in some instances can even be adapted to run on a network of communicating cores~\cite{fowler2013minimum}.  On the face of it these parallelised variants are similar to our proposal, but such schemes still need long range communication between cores and this is achieved locally by routing communications across a network of cores.  Such messages must be communicated over roughly $\log(L)$ distances and consequently also incurring order $\log(L)$ time lag.  The advantage of our proposal is two-fold.  First, we require no explicit message routing system and so our decoder could be implemented using considerably simpler cores, allowing for greater miniaturisation. Second, we have numerically benchmarked the runtime of our proposal, whereas prior proposals for parallelisations have not been simulated and so the roughly logarithmic runtime has not yet been confirmed.  These merits occur because we take a bottom-up approach which is parallel from the start, rather than attempting an \textit{ad hoc} modification of existing serial decoders.

The second class of ideas is also based on cellular automata.  It is commonly believed, that quantum error correction in two dimensions is related to the positive rates conjecture for classical systems in 1D. In analogy to the seminal work of Gacs~\cite{Gacs,gray2001} on the resolution of the positive rates conjecture, one could expect that there exists a local cellular automaton decoder with update rules that are strictly system size independent.  Harrington sketches such a program for toric code decoders~\cite{Harrington2004}, but only gives explicit details for a scheme with logarithmic overhead.  Unfortunately, in Gacs' original work, the proven noise threshold is prohibitively small ($\sim 2^{-1000}$), although the actual threshold might be considerably larger~\cite{gray2001}. Hence, the  requirements might render the proposal practically infeasible. Furthermore,  Gacs' or Harrington's update rules are more complicated than ours, again requiring more sophisticated computing hardware. Our proposals are based on physically motivated rules, have high thresholds with only logarithmic system size dependences for local parameters, and are therefore fully adequate for realistic implementations.

Our proposed 3D decoder and the analysis of the field profile is based on a working principle that is fundamentally different from all previous classes of approach.  It is the only proposal that could be implemented on a simple multi-core architecture with cores storing a single variable and mundane I/O protocols (no message routing).  The surprising result of our work is, that all the information that is required for local decoding decisions can be encoded in a physically motivated, attractive scalar field.   The propagation of information using a field also implies intrinsic robustness against small deviations in field values and updates. Indeed, preliminary results indicate that our scheme also works if the local rules are applied asynchronously, relaxing the requirement for perfect synchronous operation of the decoding unit.

We have focused on some specific instances of cellular automata decoders working against a particular noise model, but the research project opens up many new possibilities within the same paradigm.  The space of potential cellular automata is vast, small variations such as to anyon movement rules or lattice geometry could have substantial consequences on performance.  Gauss's law has proved an invaluable tool to our intuition, but it was just a guide, and departures from an electrostatic mindset could prove rewarding.  We expect our models to be naturally robust to certain types of more invasive errors, because the field  encodes global information in a smooth and local manner. Nevertheless, a more detailed study of correlated noise and errors in the measurements and in the field would be a valuable extension of the present study.  The latter two points are especially interesting as stepping stones towards extending our proposal to a passive dissipative quantum memory.  Exploring this wealth of decoder models will surely reveal many technological opportunities and a rich world of varied automata behaviors.

\section{Acknowledgements}

MJK thanks K.\ 
Michnicki for very valuable insight and motivation in the buildup to the present work. We also thank D.\ Poulin for --
upon completion of the present work -- making us aware
of the existence of unpublished results by D.\ Poulin and G.\ Duclos-Cianci, who have previously considered cellular automata based decoders in the 
same spirit as us \cite{PrivatePoulin}. A set of slides from 2011 alluding to their work are available online \cite{SlidesPoulin}. We acknowledge support from  the EU 
(SIQS, RAQUEL, COST, AQuS), the Alexander-von-Humboldt Foundation, 
the FQXi, the BMBF,
and the ERC (TAQ).

\clearpage
\appendix*
\renewcommand{\thefigure}{A.\arabic{figure}}
\renewcommand{\thetable}{A.\arabic{table}} 
\setcounter{figure}{0}
\setcounter{table}{0}
\setcounter{equation}{0}

\section*{Supplementary Material}

\subsection{Toric codes and noise models}

Any lattice that is locally two-dimensional defines a surface code.  A lattice is composed of faces, edges and vertices.  At each edge resides a qubit.  Each face $f$ is surrounded by a set of edges, which we denote $\mu(f)$. Similarly, each vertex, $v$, has a set of edges incident to the vertex, denoted $\mu '(v)$. Faces and vertices have associated operators
\begin{align}
    S_{f} &= \bigotimes_{e \in \mu(f)} Z_{e} \,, &
    S_{v} &= \bigotimes_{e \in \mu'(v)} X_{e} .
\end{align}
We consider square lattices on the torus, which are equivalent to an $L\times L$ square lattice with periodic boundary conditions.  Consequently, we can label vertices using a vector of two integers modulo $L$, denoted $\vec{v} \in \mathbb{V} := \mathbb{N}_{L}^{2}$.  A quantum state is in the code space, or ground state, when it is in the ``$+1$'' eigenspace for all face and vertex operators.  If any face (vertex) operator is measured and yields ``$-1$'', then we say an excitation, or anyon, is present at that face (vertex).

A Pauli operator acting on the code space, such as from noise, will typically produce anyons.  Any Pauli can be written as $P \propto X[E]Z[E']$, where $E$ denotes a set of edges and $X[E]=\otimes_{e \in E} X_{e}$ and similarly for $Z[E']$.  The $X[E]$ component creates face anyons, whereas $Z[E']$ creates vertex anyons.  The two problems are identical, or dual, and so herein we discuss only $X[E]$ operators.  For a $X[E]$ error, we use $q_{E}$ as the indicator function for excitations at faces.  We see $X[E]$ creates an anyon at face $f$ (and so $q_{E}(f)=1$) iff the set $E$ has an odd number of edges which are facets of $f$ (and otherwise $q_{E}(f)=0$).  For $E$ that form a path of edges, anyons appear at the endpoints of the path.  A loop, also called a cycle, has no endpoints, and so no anyons ($q_{E}=0$).  Furthermore, a non-contractible loop around the torus yields a non-trivial unitary within the code space with no accompanying anyons.  

Here we design decoders that take the anyon information from a random error $X[E]$ and determine a recovery operator $X[R]$, with 
the goal that $X[E]X[R]$ acts as the identity on the code space, e.g. when $R=E$.  We fail to decode when $X[E]X[R]=X[W]$, with $W$ being a non-contractible error chain.  Essentially, decoding consists of pairing up anyons with the closest partner.  Since we observe anyons at endpoints of error chains, this pairing process requires long range information telling us the distance between anyons.  The contribution of this work is to provide a simple local method of communicating this information between anyons.

\subsection{Error model}

Throughout we assume random noise that is uncorrelated between $X$ and $Z$ errors, such that $X[E]Z[E']$ occurs with probability $\Pr(E)\Pr(E')$ where $\Pr(E)=p^{\abs{E}}(1-p)^{L^2 - \abs{E}}$.  We compare the probability of a decoder failing, with the initial error probability $p$.  A suitable decoder will have a threshold $p_{\text{th}}$, such that for all $p<p_{\text{th}}$, the failure probability can be arbitrarily reduced by increasing $L$.  Equivalently, the expected lifetime of quantum information in the code space is preserved for arbitrarily long time, by increasing the lattice size.  The merit of the uncorrelated noise model lies in its simplicity, with $X$ and $Z$ errors as statistically independent we only need to simulate half the problem.   Furthermore, decoders with the desired threshold behavior for uncorrelated noise will also have a threshold for depolarizing, thermal and other noise models.

The spatial structure of error configurations is essential for the working principle of our decoder. In Ref.\ \cite{kovalev2013fault} it was shown that error strings with a size 
larger than $O(\log L)$ are exponentially unlikely in all quantum LDPC codes. The toric code is one of the simplest and most useful instance of a quantum LDCP code \cite{kovalev2013fault}. However, there is a constant probability for a decoder to miss-identify strings, which for our decoder would correspond to creating larger strings first before contracting them. In Theorem~3, they show that the size of objects that have to be misidentified -- so called violating $(s,m)$ sets -- is also $O(\log L)$ in the same sense as before. In Ref.\ \cite{gottesman2013overhead} it was outlined that this proof only applies to minimum weight perfect matching decoders. However, the observed logarithmic scaling of the recovery sequences in our decoder (see Fig.~\ref{Fig_Main_Results}d) has to be ascribed to similar properties of the spatial error structure.

\subsection{Numerical details}

\begin{table}[b]
\caption{\textbf{Simplified pseudo code.}}
\label{tbl:pseudo}
{\parindent0pt \bfseries\slshape\raggedright
Repeat $c$ times:

\quad Parallel for all $\bm{\xi}\in\Lambda$:

\qquad $\displaystyle \phi(\bm{\xi}) = \avg_{\langle \bm{\xi'},\bm{\xi} \rangle} \phi(\bm{\xi'}) + q(\bm{\xi})$

Parallel for all $\vec{x}\in\VV$:

\quad If $q(\vec{x})\neq0$:

\qquad With probability $0.5$:

\quad\qquad Move anyon from $\vec{x}$ to $\displaystyle\argmax_{\langle \vec{y},\vec{x} \rangle} \phi(\vec{y})$ 

If still anyons present:

\quad $\tau = \tau + 1$

\quad Go to beginning

}

\medskip
\raggedright\footnotesize
Note that for the 2D$^*$ decoder the field velocity $c$ depends on the sequence $\tau$.
\end{table}

The pseudo code for our $\phi$-automaton decoders is sketched in Table~\ref{tbl:pseudo}.
All numerical results provided in this work, were produced using a C program. The $\phi$-value at each site was stored in a 64-bit floating-point variable. However, additional tests suggest that 32-bit floating-point variables are sufficient to produce the presented results. While the algorithm is highly optimized for speed on a local parallelized architecture, especially the simulation of the 3D $\phi$-automaton on a traditional architecture can be quite expensive. To optimize the benchmark performance, every decoding process is aborted and considered a fail if the number of sequences exceeds $\tau \geq 10\cdot L$ (2D, 2D$^*$) or $\tau\geq L$ (3D). This is justified by the fact that we are looking for a decoder which has an efficient runtime. It is easy to see 
that the results on required sequences presented in Fig.~\ref{Fig_Main_Results}d are not influenced by the exit condition. Creating the threshold plot presented in 
Fig.~\ref{Fig_Main_Results}b still took approximately one month on $500$ CPUs. An open Python implementation of our decoders is available \cite{PythonPihAutomaton}.

\FloatBarrier
\subsection{Supplementary plots}
\FloatBarrier

\begin{figure}[h]
\includegraphics{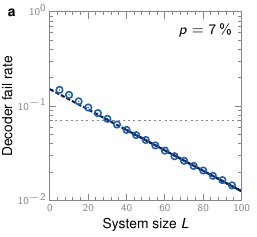}%
\includegraphics{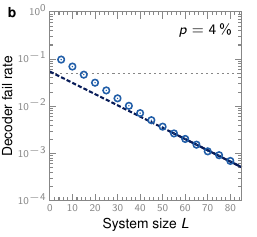}\vspace{5pt}

\includegraphics{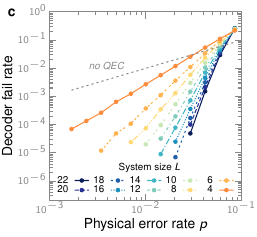}%
\includegraphics{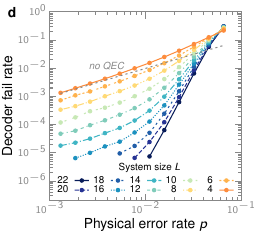}

\caption{Detailed performance analysis for the two asymptotically working decoders below threshold. Exponential suppression of fail rate in $L$ for \textbf{a} the 2D$^*$ decoder with $p=7\,\%$ and \textbf{b} the 3D decoder with $p=4\,\%$. Performance for small system sizes at low error rates for \textbf{c} the 2D$^*$ and \textbf{d} the 3D decoder. The bending of the curves in \textbf{d} deviates from the expected scaling.  Drawing any conclusions from results for system sizes $L\leq22$ is dicey since we know that we should expect finite size effects to play a role up to system size $L=55$. Every sample point is averaged over at least $10^6$ samples.}
\end{figure}

\FloatBarrier
\subsection{The matrix formalization}

In Eq.~\eqref{EQN_update}, we have
introduced the matrix formulation of Eq.~\eqref{eq:discr_poisson}
and the associated $G$ matrix.  By comparing these equations, we see 
\begin{equation}
\label{Eq_Gform}
    G_{\vec{x},\vec{y}}= 
	\begin{cases}
	(1-\eta) & \text{for } \dist(\vec{x},\vec{y})=0 , \\
    \eta/(2{D}) & \text{for }\dist(\vec{x},\vec{y})=1 , \\
    0 & \dist(\vec{x},\vec{y})>1 .
	\end{cases}    
\end{equation}
The matrix is manifestly invariant under translations, so that $G_{\vec{k}+\vec{x}, \vec{k}+\vec{y} }= G_{\vec{x},\vec{y}}$ for all $\vec{k}$, modulo the system size, 
and such matrices are said to be block circular.  Furthermore, it has the property that $\sum_{\vec{x}}G_{\vec{x},\vec{y}}=1$ and  $\sum_{\vec{y}}G_{\vec{x},\vec{y}}=1$, and so is doubly stochastic.  A consequence of doubly stochasticity is that is preserves the total field potential
\begin{equation}
\label{DoubSto}
    \sum_{\vec{x},\vec{y}}G_{\vec{x},\vec{y}} \phi_t(\vec{y}) = \sum_{\vec{y}} \phi_t(\vec{y}).
\end{equation}
It is instructive to consider the total field potential at a time $t$, which is $T_{t}=\sum_{\vec{y}}\phi_{t}(\vec{y})$.  If no anyons are present, so $\phi_{t+1}=G\phi_{t}$, 
then $T_{t+1}=T_{t}$. However, in the presence of anyons $\phi_{t+1}=G\phi_{t}+q$ and so $T_{t+1}=T_{t}+Q$, 
where $Q=\sum_{\vec{x}}q(\vec{x})$.  In this setting, the absolute field never equilibrates, though rescaling the fields resolves this problem.
Note that for practical implementations it is possible to avoid the infinity growing field by local generation of the gradient field $\bm{\nabla} \phi$. For simplicity we will restrict ourselves to the analysis of the generation of the potential $\phi$.

\subsubsection{Diagonalizing the update matrix} 

Since the update rules are translationally invariant, the associated update matrix ${G}$
 is block circulant. This entails that the Fourier transform can be employed to diagonalize this matrix and so provide solutions to many dynamical questions.  The kernel of the multi-dimensional Fourier transform $F$ has entries
\begin{equation}
	F_{\vec{j},\vec{k}} = \frac{1}{L^{D/2}} e^{-2 \pi \mathrm{i}\,  \vec{j} \cdot\vec{k}/L}.
\end{equation}
$G$ can be diagonalized as
\begin{equation}
	(F G F^\dagger)_{\vec{j},\vec{k}} = \delta_{\vec{j},\vec{k}} \sum_{\vec{l}\in \Lambda} G_{\vec{l}} e^{-2 \pi \mathrm{i}\,  \vec{j} \cdot \vec{l}/L}.
\end{equation}
What is more, 
\begin{equation}
	G_{\vec{j},\vec{k}} = G_{\vec{j}-\vec{k}} = \frac{1}{L^{D}}\sum_{\vec{l}\in \Lambda } \lambda_{\vec{l}}
	e^{2 \pi \mathrm{i}\,  \vec{l} \cdot
	(
	{\vec{j}-\vec{k}}
	)
	/L
	},
\end{equation}	
where the second expression reflects the block circulant property of $G$.
The eigenvalues of $G$ are given by
\begin{equation}
	\lambda_{\vec{k}} = \sum_{\vec{l} \in \Lambda} 
	G_{\vec{l}}
	\cos(2\pi \vec{l} \cdot\vec{k}/L ) .
\end{equation}
Using Eq.~\eqref{Eq_Gform} one immediately finds
\begin{equation}
\label{Eq_Lambda}
	\lambda_{\vec{k}} = 1-\eta + \frac{\eta}{ D} \sum_{j=1}^{ D} 
	\cos(2\pi k_j/L) 
\end{equation}
for $\vec{k}=(k_1,\dots, k_{ D})^T$. 
Furthermore, the eigenvectors are 
\begin{equation}
    \vec{v}_{\vec{k}}(\vec{x}) =e^{2 \pi \mathrm{i}\,  \vec{k} \cdot \vec{x} / L} / L^{ D/2} ,
\end{equation}
which are simply Fourier waves.

\subsubsection{Stationary field configurations}

In this section, we identify
field configurations that are stationary under field updates assuming a fixed anyon configuration.  As defined in the main text, we use $\tilde{\phi}_{t}$ to denote a rescaled field $\phi_{t}$ so that $\sum_{\vec{x}}\tilde{\phi}_{t}(\vec{x})=0$.  The renormalized fields equilibrate to a stationary value, and we now set out to estimate the equilibration time. The dynamical equation for the field is 
\begin{equation}
	\phi_{t}= G^t \phi_{0} + \sum_{m=0}^{t-1} G^{m} q. 
\end{equation}
Using $\tilde{\phi_{0}}=\phi_{0}-T_{0}$ and $\tilde{\phi_{t}}=\phi_{t}-(T_{0}+Qt)$, we have
\begin{equation}
    \tilde{\phi}_{t} = G^t \tilde{\phi}_{0} + \sum_{m=1}^{t-1}G^m q_{E}  -Qt.
\end{equation}
Let us consider the large $t$ limit.  Since $G$ is doubly stochastic, the absolute value of the eigenvalues are no greater than unity, and indeed Eq.~\eqref{Eq_Lambda} shows that only the $\vec{k}=\vec{0}$ 
subspace has a unit eigenvalue.  Consequently, $G^{t}$ tends to the projector onto the $\vec{v}_{\vec{k}=\vec{0}}$ vector. However, all rescaled fields have no $\vec{v}_{\vec{0}}$ component and so the first term vanishes. Considering the second term, and denoting by $P_{\vec{k}}=\vec{v}_{\vec{k}} \vec{v}_{\vec{k}}^{\dagger}$ 
a projector onto the eigenvector $\vec{v}_{\vec{k}}$, we have
\begin{equation}
    \sum_{m=1}^{t-1}G^m = \sum_{\vec{k}} \left( \sum_{m=1}^{t-1} \lambda^{m}_{\vec{k}} \right) P_{\vec{k}} \rightarrow \sum_{\vec{k}} (1-\lambda_{\vec{k}} )^{-1} P_{\vec{k}},
\end{equation}
where we have used the well-known limit of geometric series. So far we have
\begin{equation}
    \tilde{\phi}_{t} \rightarrow \sum_{\vec{k}} (1-\lambda)^{-1} P_{\vec{k}}q_{E} - Qt.
\end{equation}
Again using the normalization condition, we know the $\vec{k}=\vec{0}$
and constant $-Qt$ terms must cancel exactly, 
so that
\begin{equation}
    \tilde{\phi}_{t} \rightarrow \tilde{\phi}_{\infty}  =  \left( \sum_{\vec{k} \neq \vec{0}} (1-\lambda_{\vec{k}})^{-1} P_{\vec{k}} \right) q_{E}   .
\end{equation}
Using $P_{\vec{k}}=\vec{v}_{\vec{k}} \vec{v}_{\vec{k}}^{\dagger}$, we find 
\begin{equation}
    \tilde{\phi}_{\infty}(\vec{x}) = \sum_{\vec{k}   \in \Lambda \setminus \{ \vec{0} \} } a_{\vec{k}}  e^{2 \pi \mathrm{i}\,  \vec{k} \cdot \vec{x} / L },
\end{equation}
where
\begin{equation}
    a_{\vec{k}} =\frac{\sum_{\vec{x}} e^{-2 \pi \mathrm{i}\,  \vec{k} \cdot \vec{x} /L} q_{E}(\vec{x}) }{L^{ D}(1-\lambda_{\vec{k}})}.
\end{equation}
For a single anyon at the origin, the stationary solution is defined as $\varphi(\vec{x})$ and the Fourier coefficients simplify 
to $a_{\vec{k}}=(L^{ D }(1-\lambda_{\vec{k}}))^{-1}$, so that
\begin{equation}
    \varphi(\vec{x}) = \sum_{\vec{k} \neq \vec{0}}   \frac{e^{2 \pi \mathrm{i}\,  \vec{k} \cdot \vec{x} / L }}{L^{ D}(1-\lambda_{\vec{k}})} .
\end{equation}
More generally, a single anyon at cell $\vec{y}$ has a stationary field with $a_{\vec{k}}=e^{-2 \pi \mathrm{i}\,  \vec{k} \cdot \vec{y}}(L^{ D/2}(1-\lambda_{\vec{k}}))^{-1}$, so that
\begin{equation}
\label{Eq_Shifted}
    \varphi_{\vec{y}}(\vec{x})= \sum_{\vec{k} \neq \vec{0}} e^{-2 \pi \mathrm{i}\,  \vec{k} \cdot \vec{y} /L} 
    \frac{e^{2 \pi \mathrm{i}\,  \vec{k} \cdot \vec{x} / L }}{L^{ D}(1-\lambda_{\vec{k}})}.
\end{equation}
One can verify that this solution agrees with translation invariance since $\varphi_{\vec{y}}(\vec{x})=\varphi(\vec{x}-\vec{y})$. For more than a single anyon the stationary state is a linear combination of the contribution from each anyon.

\subsection{Global equilibration}

We now proceed to estimate the number of field updates that are needed for a non-equilibrium field, say $\tilde{\phi}_{0}$, to approach a stationary field $\tilde{\phi}_{\infty}$. We first observe that we can time evolve $\tilde{\phi}_{\infty}$ by $t$ time steps, and by virtue of being a stationary state, we have 
\begin{equation}
    \tilde{\phi}_{\infty} = G^t \tilde{\phi}_{\infty} + \sum_{m=1}^{t-1}G^m q  -Qt.
\end{equation}
Taking the difference of $\tilde{\phi}_{\infty}$ and $\tilde{\phi}_{t}$, only the first terms remain
\begin{equation}
\label{Eq_ReExpressed}
    \tilde{\phi}_{t} - \tilde{\phi}_{\infty} = G^t(\tilde{\phi}_{0} - \tilde{\phi}_{\infty}).
\end{equation}
By the H\"{o}lder inequality, for a single charge at the origin,
\begin{eqnarray}
\label{EQ_General_Convergence}
    \norm{ \tilde{\phi_{t}} - \tilde{\phi_{\infty}} }_{2} & \leq &  \lambda_{\text{max}}^t \norm{ \tilde{\phi_{0}} - \tilde{\phi_{\infty}} }_{2},
\end{eqnarray}
where fields are measured in the vector 2-norm $\norm{\vec{u}}_2^2=\vec{u}\cdot \vec{u}$, 
and $\lambda_{\text{max}}=\max_{\vec{k} \neq \vec{0}}  \abs{ \lambda_{\vec{k}} }$.  Again, the $\vec{k}=\vec{0}$ eigenvalue can be neglected since the fields are normalized to have no overlap with this eigenvector.  Provided $\eta \leq 1/2$, we have that $\abs{ \lambda_{\vec{k}} } = \lambda_{\vec{k}}$,  and we find
\begin{equation}
	\lambda_{\text{max}} =	 1-\eta + \frac{\eta}{ D}\biggl( \cos (2\pi/L) + ( D -1)\biggr) ,
\end{equation}
and using $\cos(x) \leq 1-x^2/4$ for $x\in[0,2]$,
we find
\begin{equation}
	\lambda_{\text{max}} \leq	1 - \frac{ \eta  \pi^2 }{D L^2 }
\end{equation}
for $L\geq 4$.
Furthermore, 
\begin{equation}
	1-x \leq e^{-x} ,
\end{equation}
for all $1\geq x\geq 0$,
 so $\lambda_{\text{max}}\leq e^{-\eta \pi^2 / ( D L^2)}$.  Combined with Eq.~\eqref{EQ_General_Convergence}
 gives the result  
\begin{equation}
    \norm{ \tilde{\phi}_{t} - \tilde{\phi_{\infty}} }_{2}  \leq  e^{-(\eta \pi^2 / D)t/L^2} \norm{ \tilde{\phi}_{0} - \tilde{\phi_{\infty}} }_{2}.
\end{equation}
While the equilibration is exponentially fast in $t$ for constant lattice size, this bound tells us little 
in the large lattice limit as $\lambda_{\text{max}}\rightarrow 1$.  This is intuitive since after an event of anyon movement the field can only change within a light cone of the event.  
Consequently, 
for any period of time ($t<L/2$) there is always a distance at which the field has not equilibrated.  However, we are interested in the local equilibration of a cluster of anyons in a small region, and contained within a larger lattice. This finer estimate is presented next.

\subsection{Convergence after anyon steps}

In this section, we consider the point-wise convergence of the field, following an anyonic hopping from $\vec{x}=\vec{0}$ to $\vec{x}=-\vec{e}$.
The initial distribution is determined by $\tilde{\phi}_{t=0}(\vec{x})=\varphi(\vec{x})$ for $\vec{x}\in \Lambda$, then at sufficiently late times $\tilde{\phi}_{\infty}(\vec{x})=\varphi(\vec{x}+\vec{e})$.
Recall that
\begin{equation}
	\tilde{\phi}_{t=0}(\vec{x}) = 
	\sum_{\vec{k}  \neq \vec{0} }
	\frac{ e^{2 \pi \mathrm{i}\,  \vec{x} \cdot \vec{k}/L}}{L^{D}(1-\lambda_{\vec{k}})}.
\end{equation}
The final distribution after a shift by one position (see Eq.~\eqref{Eq_Shifted}) is
\begin{equation}
	\tilde{\phi}_\infty (\vec{x})= 
	\sum_{\vec{k} \neq \vec{0}}
	\frac{ e^{2 \pi \mathrm{i}\,  (\vec{x} + \vec{e}) \cdot \vec{k}/L} }{L^D(1-\lambda_{\vec{k}})}.
\end{equation}
Therefore, 
\begin{equation}
	\tilde{\phi}_{0}(\vec{x})-\tilde{\phi}_{\infty}(\vec{x})  =  
	\sum_{\vec{k}  \neq \vec{0}}
	 \frac{e^{2 \pi \mathrm{i}\,  \vec{x}  \cdot \vec{k}/L}(1- e^{2 \pi \mathrm{i}\,  \vec{e}  \cdot  \vec{k}/L})}{L^{D}(1-\lambda_{\vec{k}})}. 
\end{equation}
Using Eq.~\eqref{Eq_ReExpressed}, 
we find that at time $t$
\begin{equation}
\label{Eq_Delta}
	\Delta_{t}(\vec{x})  =  	\sum_{\vec{k}  \neq \vec{0}}
	 \frac{\lambda_{\vec{k}}^t e^{2 \pi \mathrm{i}\,  \vec{x}  \cdot \vec{k}/L}(1- e^{2 \pi \mathrm{i}\,  \vec{e}  \cdot  \vec{k}/L})}{L^{D}(1-\lambda_{\vec{k}})} ,
\end{equation}
where $\Delta_{t}(\vec{x}):=\tilde{\phi}_{t}(\vec{x}) - \tilde{\phi}_\infty (\vec{x})$. We know the left hand side is real and so may take the real part of the right, and using the triangle inequality
\begin{equation}
\label{Eq_Delta_so_far}
	\Delta_{t}(\vec{x})  \leq  	\sum_{\vec{k}  \neq \vec{0}}
 \frac{ \abs{ \lambda_{\vec{k}} }^t A_{\vec{k}} }{ L^{D} B_{\vec{k}} },
\end{equation}
where we have collected many details into $A_{\vec{k}}$ and $B_{\vec{k}}$,
\begin{eqnarray}
    A_{\vec{k}} &:=&  \abs{ \re [e^{2 \pi \mathrm{i}\,  \vec{x}  \cdot \vec{k}/L}(1- e^{2 \pi \mathrm{i}\,  \vec{e}  \cdot  \vec{k}/L}) ] } , \\ \nonumber
    B_{\vec{k}}&:=&    1-\lambda_{\vec{k}}.
\end{eqnarray}
The next few steps of the derivation just concern these quantities, and culminate in the bound
\begin{equation}
\label{Cxk_upper_bound}
      \frac{A_{\vec{k}}}{B_{\vec{k}}} \leq \frac{ D \norm{ 2\vec{x} + \vec{e} }_{1} }{2 \eta} ,
\end{equation}
where  $\norm{ \vec{x} }_{1}=\sum_{j} \abs{ x_{j} }$.
	
First, we recast the numerator $A_{\vec{k}}$,
\begin{eqnarray}
\label{EQ_long_trig_inequal}
A_{\vec{k}}	&=& \Abs{ \cos\left(2\frac{\pi}{L} \vec{x} \cdot \vec{k} \right) - \cos\left(2\frac{\pi}{L} (\vec{e}+\vec{x}) \cdot \vec{k} \right) }  , \nonumber \\
	&=& 
	\Abs{ 2 \sin\left( 
	\frac{\pi}{L} \vec{k}\cdot (2\vec{x} + \vec{e}  )
	\right) \sin\left(\frac{\pi}{L } \vec{k}\cdot \vec{e}\right) }.
\end{eqnarray}
Before proceeding we demonstrate the following trigonometric inequality:   $\abs{ \sin(u n) } \leq \abs{ \sin(u) } n$ if $n$ is a positive integer. When $n=0$ both sides are zero so it clearly holds.   For positive integers, we write $un=u(n-1)+ u$ to get\begin{equation*}
   \abs{ \sin(un) } = \abs{ \sin(u(n-1))\cos(u(n-1))+\sin(u)\cos(u) }.
\end{equation*}
Using the triangle inequality and $\abs{ \cos(x) } \leq 1$, we get
\begin{equation*}
   \abs{ \sin(un) } \leq \abs{ \sin(u(n-1)) } + \abs{ \sin(u) }.
\end{equation*}
This recursive relation and the $n=0$ case shows $\abs{ \sin(u n) } \leq \abs{ \sin(u) } n$ for positive integer $n$.  We use this result on the first factor of Eq.~\eqref{EQ_long_trig_inequal}. Setting
\begin{align}
    u &= \frac{\pi \abs{ \vec{k} \cdot(2 \vec{x}+\vec{e})} }{L \norm{ 2 \vec{x}+\vec{e} }_{1}} , &
    n &= \norm{2 \vec{x}+\vec{e}}_{1} ,
\end{align}
yields
\begin{multline}
  \Abs{  \sin \left( \frac{\pi}{L}  \vec{k} \cdot (2\vec{x} + \vec{e} ) \right) }  \leq\\\leq \sin \left(  \frac{\pi \abs{ \vec{k} \cdot(2 \vec{x}+\vec{e}) } }{L \norm{2 \vec{x}+\vec{e} }_{1}} \right) \norm{ 2 \vec{x}+ \vec{e} }_{1} \,.
\end{multline}
By applying the H\"{o}lder inequality, one finds 
$\abs{ \vec{k} \cdot (2\vec{x} + \vec{e} ) } \leq \norm{ \vec{k} }_{\infty}  \norm{ 2\vec{x} + \vec{e} }_{1}$ where $\norm{ \vec{x} }_{\infty}:=\text{max}_{j} \abs{ x_{j} }$.  Therefore,
\begin{equation}
\label{Eq_U_inequality}
    u \leq  \pi \norm{ \vec{k} }_{\infty} /L .
\end{equation}
Since $\vec{k}  \in \Lambda $, $0 \leq k_{j} \leq L$ and so one might infer $\norm{ \vec{k} }_{\infty}\leq L$. However, 
we are only interested in sums of periodic functions over $\vec k$, where $\vec k$ runs over integer multiples of the functions period. This allows to choose a
coordinate system where 
\begin{equation}
\label{eq:k-coordinates}
	k_j = -\tfrac L2, -\tfrac L2+1,\dots,\tfrac L2-1,\tfrac L2 \,,
\end{equation}
and so $\norm{ \vec{k} }_{\infty}\leq L/2$.   This entails $0 < u \leq \pi /2$, which is important because $\sin(u)$ is monotonic on this interval. Monotonicity allows us to upper bound
$\sin(u)$ by $\sin(v)$ whenever $u \leq v\leq \pi/2$, and using $v$ as the RHS of Eq.~\eqref{Eq_U_inequality} we deduce
\begin{equation}
      \Abs{  \sin \left( \frac{\pi}{L}  \vec{k} \cdot (2\vec{x} + \vec{e} ) \right) }  \leq \sin \left(\frac{\pi}{L} \norm{ \vec{k} }_{\infty} \right)  \norm{2 \vec{x}+ \vec{e} }_{1}.
\end{equation}
Similarly the second $\sin$ factor of Eq.~\eqref{EQ_long_trig_inequal} satisfies $\abs{ \sin( \pi \vec{k} \cdot \vec{e} / L) } \leq \abs{ \sin( \pi \norm{ \vec{k} }_{\infty} \norm{ \vec{e} }_{1} / L) } $, 
and since $\vec{e}$ is a 
unit vector, $\abs{ \sin( \pi \vec{k}\cdot\vec{e} / L) } \leq \abs{ \sin( \pi \norm{ \vec{k} }_{\infty} /L) } $.  Combining these upper bounds with Eq.~\eqref{EQ_long_trig_inequal}, we find
\begin{eqnarray}
A_{\vec{k}}	&\leq& 2 \sin^2( \pi \norm{ \vec{k} }_{\infty} / L ) \norm{ 2\vec{x}+ \vec{e} }_{1}   ,\end{eqnarray}
which completes our simplification of the numerator $A_{\vec{k}}$.
Next we turn to the denominator $B_{\vec{k}}$, which is simply
\begin{equation}
    B_{\vec{k}}=1-\lambda_{\vec{k}} = \eta - \frac{\eta}{D} \sum_{j=1}^{D}\cos
    \left(
    2 \pi k_j /L
    \right).
\end{equation}
Clearly, $1-2\sin^2(x) = \cos(2x)$  for all $x$, so that
\begin{equation}
   B_{\vec{k}} =  \frac{2 \eta}{D} \sum_{j=1}^{D}\sin^2
    \left(
    \pi k_j /L
    \right).
\end{equation}
Bringing numerator and denominator together and rearranging we finally have
\begin{equation}
	\frac{A_{\vec{k}}}{B_{\vec{k}}} \leq \left( \frac{\sin^2 ( \pi \norm{ \vec{k} }_{\infty} / L )}{  \sum_{j} \sin^2(\pi \abs{k_j} / L ) } \right) \left( \frac{D  \norm{ 2 \vec{x} +\vec{e} }_{1}}{\eta} \right).
\end{equation}
Lastly we note the first bracket is upper bounded by unity since the numerator is simply one of the positive terms in the denominator, and this proves Eq.~\eqref{Cxk_upper_bound}.
Putting this into Eq.~\eqref{Eq_Delta_so_far} we arrive at
\begin{equation}
\label{Eq_Delta_Before_shells}
	\abs{ \Delta_{t}(\vec{x}) }
	\leq  \frac{D \norm{ 2\vec{x}+ \vec{e} }_1}{\eta L^{D}} 
	\sum_{\vec{k}  \neq \vec{0}} \abs{ \lambda_\vec{k} }^t.
\end{equation}
We proceed by dividing the eigenvalues into shells of equal $\norm{ \vec{k} }_{\infty}$. We denote with $O_{l}$ the number of elements in $\vec{k}$-space shell for which $\norm{ \vec{k} }_{\infty}=l$
for $l=1,\dots, \tfrac{L-1}2$, where we have used again the $\vec k$-coordinates defined in Eq.~\eqref{eq:k-coordinates}. Here, we have assumed for simplicity that $L$ is odd, with even $L$ admitting a similar treatment. Let us assume ${D}\geq 2$.
Clearly, we have 
\begin{equation}
	O_{l} =  (2l+1)^{D}  - (2l-1)^{D} 
	\leq 
	2^{2D-1}
	l^{{D}-1} .
\end{equation}
We can now upper bound 
\begin{eqnarray}
	V_l &:=& \sup\left\{ 
	\abs{ \lambda_{\vec{k}} }: \norm{\vec{k}}_\infty = l \right\}  \\
	&=& \lambda_{(l,0,\dots, 0)} \nonumber \\ \nonumber
	  &=& (1-\eta)+\frac{\eta}{D}\left[ \cos ( 2 \pi l / L ) + D-1 \right],
\end{eqnarray}
where we have again assumed $\eta \leq 1/2$ to ensure that $\abs{ \lambda_{\vec{k}} }=\lambda_{\vec{k}} $.  Making use of $\cos(x)\leq 1-x^2/4$ for $x\in[-\pi/2,\pi/2]$, one obtains
\begin{equation}
	V_l  \leq 1- 
	\frac{\pi^2 \eta}{ D} \frac{l^2}{L^2}.
\end{equation}
This gives
\begin{eqnarray}
	\sum_{\vec{k}\neq 0} \abs{ \lambda_{\vec{k}} }^t
	&\leq & \sum_{l=1}^{(L-1)/2} O_{l} V_{l}^{t} \\ \nonumber
	&\leq & 2^{2D-1} \sum_{l=1}^{(L-1)/2} l^{{D}-1} \left(
	 1- 	\frac{\pi^2 \eta}{D} \frac{l^2}{L^2}	\right)^t.\nonumber
\end{eqnarray}
Plugging this into Eq.~\eqref{Eq_Delta_Before_shells} and collecting constants, gives
\begin{equation}
    \abs{ \Delta_{t}(\vec{x}) } \leq \frac{\alpha}{L} \sum_{x=\frac{1}{L},\frac{2}{L},\dots, \frac{L-1}{2L}} x^{D-1}(1-\beta x^2)^t
\end{equation}
where 
\begin{align}
    \alpha & :=  \frac{D \norm{ 2 \vec{x} + \vec{e} }_1 2^{2D-1}}{\eta} , &
    \beta & :=     \frac{\pi^2 \eta}{D} , &
    x & :=  l/L.
\end{align}
We surely have that
\begin{eqnarray}
	\abs{ \Delta_{t}(\vec{x}) }=
	\abs{ \tilde{\phi}_{t}(\vec{x}) - \tilde{\phi}_\infty (\vec{x}) }
	&\leq & \epsilon,
\end{eqnarray}	
if 
\begin{equation}
	x^{{D}-1} \left(
	1- \beta x^2 \right)^t\leq  \epsilon / \alpha ,
\end{equation}
for all $x\in [0,1/2]$.  This is satisfied provided
\begin{eqnarray}
	t \geq t_{\rm min}:=\max_{x\in [0,1/2]}\frac{\log\left( \epsilon x^{1-D} / \alpha 
	\right)}{\log\left(
	1- \beta x^2 \right)}.
\end{eqnarray}
We can now use that
\begin{equation}
	-\log(1- \beta x^2)^{-1} \leq \frac{1}{\beta x^2},
\end{equation}
for $\beta x^2 \in[0,1/2]$ (which can be inferred from $\eta\leq 1/2$, $D \geq 2$, and $x \leq 1/2$), to bound
\begin{equation}
\label{eq_Tmin}
	t_{\rm min} \leq \max_{x\in [0,1]} \frac{\log\left(   \alpha x^{D-1} 
/ \epsilon   \right) }{\beta x^2} .
\end{equation}
The extremum of 
\begin{equation}
	f(y)= \log(\alpha y^{{D}-1})/y^2 
\end{equation}
can be found by differentiation to be at $y_{\text{max}}=e^{1/2} \alpha^{-1/({D}-1)}$ taking the functional value $f(y_{\text{max}})=(D-1)\alpha^{2/(D-1)}/2e  $. In Eq.~\eqref{eq_Tmin} we have this form with $a=\alpha / \epsilon $ and so
\begin{equation}
	t_\text{min} \leq  \frac{D-1}{2\beta e}\left( \frac{\alpha}{\epsilon}\right)^{\frac{2}{D-1}}  .
\end{equation}
Inserting the values of $\alpha$ and $\beta$ we get
\begin{equation}
	t_{\rm min}\leq \chi(\vec{x}) \epsilon^{\frac{2}{1-D}}
\end{equation}
where
\begin{multline}
\chi(\vec{x})
:=
\frac{8}{e \pi^2} (d-1) \left(\frac{d}{\eta}\right)^{\frac{2}{D-1}+1} \left(  2^{2d-1} \Norm{2\vec x - \vec e}_1 \right)^{\frac{2}{D-1}}
\\\leq
\frac{d-1}{2e \pi^2} \left(\frac{d}{\eta}\right)^{\frac{2}{d-1}+1} \left[4^d \left( \Norm{\vec x}_1+ \tfrac12 \right) \right]^{\frac{2}{d-1}} \,.
\end{multline}
This provides Eq.~\eqref{tmin_rough} with an explicit value of $\chi(\vec{x})$.  Setting $\vec{x}=0$ and $\eta=1/2$, we find $\chi(\vec{0}) < 77$ and $\chi(\vec{0}) <43$ for dimensions $D=2,3$,
 respectively.
 For general $\vec{x}$, we can obtain a neater upper bound on $\chi(\vec{x})$, so that
\begin{equation}
\label{Chi_approx}
\chi(\vec x)\leq
\frac{1}{e \pi^2} \cdot
\begin{cases}
1024 \cdot \left(\Norm{\vec x}_1+\frac12\right)^2  \eta^{-3} & D = 2 \\
576\cdot \left(\Norm{\vec x}_1 + \frac12 \right) \eta^{-2} & D = 3 \\
\end{cases}\,,
\end{equation}
This be achieved using the triangle inequality, so $\norm{ 2 \vec{x} + \vec{e} }_{1} \leq 2 \norm{ \vec{x} }_{1}+ \norm{ \vec{e} }_{1}=2 \norm{ \vec{x} }_{1}+1$.

\end{document}